\documentstyle[12pt,aasms4]{article}

\begin{document}

\title{Breaking Degeneracy of Dark Matter Models by the Scale-Scale 
Correlations of Galaxies}

\author{Longlong Feng\footnote{fengll@physics.arizona.edu}, 
Zu-Gan Deng\footnote{dzg@qso.bao.ac.cn},
and Li-Zhi Fang\footnote{fanglz@physics.arizona.edu}
}
\affil{$^1$Center for Astrophysics, University of Science and Technology of 
China, \\
National astronomical Observatory, Chinese
Academy of Sciences\\
Hefei, Anhui 230026, China\\}
\affil{$^2$ Graduate School, USTC, Chinese Academy of Sciences, P. O. 
Box 3908 \\
Beijing 100039, China\\
Beijing Astrophysics Center, National astronomical Observatory, \\
Chinese Academy of Sciences, Beijing 100012, China\\}
\affil{$^3$Department of Physics, University of Arizona, Tucson, AZ 85721}

\begin{abstract}

Recently, scale-scale correlations have been detected in the distributions
of quasar's Ly$\alpha$ absorption lines and the maps of cosmic temperature
fluctuations. In this paper, we investigate the scale-scale correlations in
galaxy distributions. Using samples of mass field given by N-body
simulation, we first show that the scale-scale correlation of 2-D and
3-D mass distributions at present day is capable of breaking the degeneracy
between the SCDM (standard cold dark matter model) and
OCDM (open CDM model) or LCDM (flat CDM model), and even show the difference
between the OCDM and LCDM. 
Using biased galaxy samples produced in an appropriate bias model,  
we show that the scale-scale correlation of
galaxy distribution at zero redshift is still a powerful tool to break the
degeneracy in the parameter space, including both cosmological and
biasing parameters. 
We analyze the scale-scale correlations of the APM
bright galaxy catalog, and compare it with the mock catalog in the SCDM,
OCDM and LCDM models. We find that all the spectra of local and
non-local scale-scale correlations predicted by the OCDM model are
in excellent agreement with those of the APM-BGC sample, while the
SCDM and LCDM mock catalog appear to have somewhat weaker scale-scale
correlations than the observation.

\end{abstract}

\bigskip


\newpage

\section{Introduction}

A problem in current study of the large scale structures of the universe is 
the so called {\it degeneracy} among viable models. That is, structure
formation models with different cosmological parameters are able to
reproduce the same features as the observed large-scale structure. For
instance, in the cold dark matter (CDM) cosmogony, the standard CDM model
(SCDM) predicts about the same abundance and two-point correlation functions
of the present ($z=0$) galaxy clusters as the open CDM model (OCDM) and the
low density flat model (LCDM). This is sometimes called the degeneracy in
parameter space of $\Omega$ and $\sigma_8$, where $\Omega$ is the present
value of the cosmological density parameter, and $\sigma_8$ the normalization
of the power spectrum on a 8 $h^{-1}$ Mpc scale. Considering the
redshift evolution of clusters, the degeneracy between the SCDM and OCDM
(or LCDM) is broken (e.g. Jing \& Fang 1994, Bahcall, Fan \& Cen 1997.)
However, the models of OCDM and LCDM remain in degeneracy for the second
and lower order statistical features of clusters till redshifts as high as
$z \sim 0.5$.  

In terms of the clustering of galaxies, model-degeneracy is more common.
For instance, the mock catalog of SDSS and 2dF galaxy redshift surveys
were generated by a bias sampling of the simulated mass fields.
The parameters used for the bias sampling were selected by fitting the
two-point correlation function of the bias sampled galaxies with observations
(Cole et.al, 1998.) Thus, a majority of the dark matter models are found to
be able to produce mock catalog which is basically in agreement with the
observed power spectrum and two-point correlation function of galaxies if
the biasing parameters are appropriate selected. Namely, to a first 
approximation, all the models considered are degenerate. This is a degenerate 
in the space of both dark matter and biasing parameters.

It is largely believed that the non-linear behaviors of cosmic clustering
would be useful for the degeneracy breaking. In this paper, we will investigate
the model-degeneracy breaking by a non-linear measure, the local scale-scale
correlations. The hierarchical scenario of structure formation is
characterized by a rule that determine how are the small scale massive halos
related to their parent halos on large scale. The relation between parent
and daughter halos leads to a local scale-scale correlation. Different models 
are characterized by different relations between massive halos on different
scales. Therefore, the local scale-scale correlation is sensitive to models. 
This motivated us to study the possibility of breaking the model-degeneracy 
by the local scale-scale correlations.

Moreover, the local scale-scale correlations have been detected in 1-D 
samples like Ly-$\alpha$ forests (Pando et al. 1998), and 2-D samples from 
COBE-DMR cosmic temperature map on angular scales of 10 - 20 degrees (Pando, 
Vills-Baude \& Fang 1998). Namely, the current data of large scale
structures are ready for detecting scale-scale correlations. In this paper,
we show that the scale-scale correlations of the APM-BGC galaxies can
provide meaningful information for degeneracy breaking of cosmological
models.

The outline of this paper is as follows. Based on the discrete wavelet
transform (DWT), \S 2 gives a brief description of the basic idea and
algorithm for the scale-scale correlations. In \S 3 the scale-scale
correlations are computed in simulated samples for three typical CDM models.
The scale-scale correlation of underlying dark matter and biased galaxies
in real space as well as redshift space are investigated in detail.
The comparison of the scale-scale correlations between the APM-BGC
galaxies and the model predictions are presented in \S 4. Finally in \S 5,
we briefly summarize the conclusions drawn from our calculations.

\section{Basic formulae of scale-scale correlations}

In the hierarchical clustering scenario, the evolution of massive halos is
prescribed by an evolutionary tree of halos, such as smaller halos merging 
into a bigger one. This process leads to correlations between 
density perturbations on different scales, but spatially localized. The
correlation can easily be described in phase space, i.e., the 
position-wavevector space. For phase space analysis, a density field
$\rho({\bf r})$ is decomposed into perturbations at phase space points
$({\bf k}, {\bf x})$, where ${\bf k}$ and ${\bf x}$ denote, respectively, the 
wavevector and position of a volume element $d^3 x d^3 k \simeq 1$ in phase
space. The localized scale-scale correlation characterizes the correlations 
between the density perturbations at two phase space points with the same 
spatial coordinate ${\bf x}$, but different scale coordinates ${\bf k_1}$ 
and ${\bf k_2}$.

One way to realize the phase space decomposition is by the discrete wavelet 
transform (DWT) (Fang \& Thews 1998). With the DWT decomposition, a 3-D 
density field  $\rho({\bf x})$, ${\bf x}=(x_1,x_2,x_3)$, in the range 
$0<x_i<L_i, \{i=1,2,3\}$ can be expanded as
\begin{equation}
\rho({\bf x})=\bar{\rho}+\sum_{\bf j}^{\infty}\sum_{\bf l=0}^{2^{\bf j}-1}
\tilde{\epsilon}_{\bf j,l}\psi_{\bf j,l}({\bf x})
\end{equation}
$\psi_{\bf j,l}({\bf x})$ are the bases of the DWT, ${\bf j}= (j_1,j_2,j_3)$, 
${\bf l}=(l_1,l_2,l_3)$ and $j_i=0,1,..$, $l_i=0,1,...2^{j_i}-1$. 
Because these bases are orthogonal and complete, the wavelet function
coefficient (WFC), $\tilde{\epsilon}_{\bf j,l}$, is 
computed by
\begin{equation}
\tilde{\epsilon}_{\bf j,l}= 
\int \rho({\bf x}) \psi_{\bf j,l}({\bf x})d{\bf x}.
\end{equation}
 From the construction of wavelet bases, we know that indices 
${\bf j}$ and ${\bf l}$ corresponds, respectively, to the scale 
$(L_i/ 2^{j_i}, i=1,2,3)$ and position ${\bf x}=\{L_il_i/2^{j_i}, i=1,2,3\}$.  
Sometimes we use wavevector ${\bf k}$, which defined as 
${\bf k}/2\pi = \{ 2^{j_i}/L_i, i=1,2,3\}$.  
Therefore, $\tilde{\epsilon}_{\bf j,l}$ represents the amplitude of the 
density fluctuation localized at point $({\bf k},{\bf x})$ of phase space. 

In general, density fields in cosmology are assumed to be ergodic, i.e., the
average over an ensemble is equivalent to the spatial average in one
realization. Thus, the ensemble average of perturbations on scale
${\bf j}$ is
\begin{equation}
\langle \tilde{\epsilon}_{\bf j} \rangle 
=\bar{\sum_{\bf j}} \tilde{\epsilon}_{\bf j,l}=0.
\end{equation}
where the normalized summation notation $\bar{\sum_{\bf j}} \equiv
\prod_{i=1}^{3}\frac{1}{2^{j_i}}\sum_{l_i=0}^{2^{j_i}-1}$ was used.
The last step of eq.(3) is due to the fact that the bases of wavelet are 
admissible, or 
$\int\psi({\bf x}) d{\bf x}=0$. Thus, the variance or the power spectrum 
with respect to the DWT decomposition is determined by (Pando \& Fang, 1998)
\begin{equation}
P_{DWT}({\bf j}) = \frac{1}{2^{j_1+j_2+j_3}}
\langle\tilde{\epsilon}_{\bf j}
\tilde{\epsilon}_{\bf j}\rangle = 
\bar{\sum_{\bf j}}
\tilde{\epsilon}^2_{\bf j,l}.
\end{equation}

The local correlation between perturbations on scales of
${\bf j}=(j_1,j_2,j_3)$ and ${\bf j+1}=(j_1+1,j_2+1,j_3+1)$ can be
measured by 
\begin{equation}
\langle \tilde{\epsilon}_{\bf j}\tilde{\epsilon}_{{\bf j}+1}\rangle
=\bar{\sum_{\bf j+1}}
\tilde{\epsilon}_{\bf j,[l/2]} 
\tilde{\epsilon}_{{\bf j}+1,{\bf l}},
\end{equation}
where the [\hspace{2mm}]'s denote the integer part of the quantity. Because 
$L_1l/2^j = L_12l/2^{j+1}$, the position $l_1$ at the scale $j_1$ is the
same as the positions $2l_1$ and $2l_1+1$ at the scale $j_1+1$. Therefore,
eq.(5) measures the correlation between scales ${\bf j}$ and ${\bf j}+1$
at the {\it same} physical point ${\bf x}$, i.e. it is localized scale-scale
correlation. If perturbations on scales ${\bf j}$ and ${\bf j}+1$ are
independent, we have
$\langle\tilde{\epsilon}_{\bf j}\tilde{\epsilon}_{{\bf j}+1}\rangle
\propto \langle\tilde{\epsilon}_{\bf j}\rangle
 \langle \tilde{\epsilon}_{\bf j+1}\rangle =0$.
It is interested to point out that the statistic of eq.(5) is of second
order. 

The statistical properties of $\tilde{\epsilon}_{{\bf j},{\bf l}}$ with 
respect to ${\bf l}$ generally are independent from the statistics on 
${\bf j}$. For instance one can have distributions for which 
$\tilde{\epsilon}_{{\bf j},{\bf l}}$ is Gaussian in
its one-point distribution with respect to ${\bf l}$ while scale-scale
correlated in terms of ${\bf j}$ (Greiner, Lipa \& Carruthers, 1995).
A simple example is as follows. Considering a
distribution of $\tilde{\epsilon}_{{\bf j}_1,{\bf l}}$ which is Gaussian in
${\bf l}$ at a given  scale ${\bf j}_1$. Namely, higher order correlations
of $\tilde{\epsilon}_{{\bf j}_1,{\bf l}}$ are zero at this scale. Let's 
consider the perturbation $\tilde{\epsilon}_{{\bf j}_2,{\bf l}}$ at 
scale ${\bf j_2}$. If at a given physical point ${\bf x}$, 
$\tilde{\epsilon}_{{\bf j}_2,{\bf l}}$ is always proportional 
to $\tilde{\epsilon}_{{\bf j}_1,{\bf l}}$ at the same point, the distribution
of $\tilde{\epsilon}_{{\bf j}_2,{\bf l}}$ is also Gaussian in the ${\bf l}$
distribution. However, the perturbations on scales ${\bf j_1}$ and ${\bf j_2}$
are strongly correlated. Obviously such localized scale-scale correlations 
can only be detected by a scale-space decomposition.

Instead of eq.(5), we will employ a normalized scale-scale correlation 
defined as
\begin{equation}
C^{p,p}_{{\bf j}}=
\frac{\displaystyle{\bar{\sum}_{{\bf j}+1}}
\tilde{\epsilon}^p_{\bf j,[l/2]}\tilde{\epsilon}^p_{{\bf j}+1,{\bf l}}}
{\displaystyle{\bar{\sum}_{\bf j}\tilde{\epsilon}^p_{\bf j,[l/2]}
\bar{\sum}_{{\bf j}+1}\tilde{\epsilon}^p_{{\bf j}+1,{\bf l}}}},
\end{equation}
where $p$ is an integer. In the case of odd $p$, the denominator 
would be zero [eq.(3)], and therefore, we will use even $p$ only. 
The statistics $C^{p,p}_{\bf j+1}$ 
measures the $p$-order correlations between local density perturbations
with wavevectors ${\bf k}/2\pi = (2^{j_1}/L_1, 2^{j_2}/L_2, 2^{j_3}/L_3)$ and
${\bf k'}/2\pi =(2^{j_1+1}/L_1, 2^{j_2+1}/L_2, 2^{j_3+1}/L_3)$. 
In the case of $p=2$, $C^{2,2}_{\bf j+1}$ is actually the correlation between 
the DWT spectra [eq.(4)] on scales ${\bf j}$ and ${{\bf j}+1}$, i.e. it is
a local spectrum-spectrum correlations.

Generally, we can measure the $p$-order correlation between perturbations 
at two phase space points $({\bf j}, {\bf l})$ and 
$({\bf j}+1, {\bf l + \Delta l})$ 
by
\begin{equation}
C^{p,p}_{{\bf j}, \Delta{\bf l}}=
\frac{\displaystyle{\bar{\sum}_{{\bf j}+1}}
\tilde{\epsilon}^p_{\bf j,[l/2]}
\tilde{\epsilon}^p_{{\bf j}+1,{\bf l+\Delta l}}}
{\displaystyle{\bar{\sum}_{\bf j}\tilde{\epsilon}^p_{\bf j,[l/2]}
\bar{\sum}_{{\bf j}+1}\tilde{\epsilon}^p_{{\bf j}+1,{\bf l}}}}
\end{equation}
For $p=2$ and $\Delta l \neq 0$, $C^{2,2}_{{\bf j}+1, \Delta{\bf l}}$ 
describes non-local spectrum-spectrum correlations.
 
Obviously, $C_{\bf j}^{p,p}=1$ and $C^{p,p}_{{\bf j}+1, \Delta{\bf l}}=1$
for Gaussian fields. The inequality
$C_{\bf j}^{p,p} >1$ and $C^{p,p}_{{\bf j}+1, \Delta{\bf l}}>1$
correspond to a positive scale-scale correlation, and
$C_{\bf j}^{p,p} <1$ and $C^{p,p}_{{\bf j}+1, \Delta{\bf l}}<1$
to the negative case. Despite the statistics of eqs. (6) and
(7) being higher order for $p \geq 2$, they are essentially the two-point
statistics in phase space.

In this paper, we will consider only diagonal components of scale-scale
correlation, i.e., $L_1=L_2=L_3$, $j_1=j_2=j_3=j$ and
$\Delta l_1=\Delta l_2=\Delta l_3= \Delta l$.  

\section{Scale-scale correlation of simulation samples}

\subsection{Samples of N-body simulation}

To demonstrate the scale-scale correlations as a tool of dengeneracy breaking,
we first calculate the scale-scale correlation of mass fields given by
N-body simulation. We used the modified version of AP3M code (Couchman, 1991). 
The cosmological models considered here are standard cold dark matter
model (SCDM), low density open model (OCDM) and low density flat model 
(LCDM). The model parameters $(\Omega_0, \Omega_{\lambda}, \Gamma, \sigma_8)$
are taken to be $(1.0,0.0,0.5,0.55)$, $(0.3,0.0,0.25,0.85)$ and 
$(0.3,0.7,0.25,0.85)$ for the SCDM, OCDM and  LCDM, respectively.

The simulations were performed by evolving $128^3$ particles in a periodic 
cubical box of side $256h^{-1}$Mpc. The soft parameter is taken to be
15\% of grid spacing, i.e. $300h^{-1}$ kpc, which keep unchanged in the
comoving coordinates, and the force resolution approximates 100 $h^{-1}$ kpc
for a Plummer force law correspondingly. The initial perturbations
were generated by the Zeldovich approximation applying to a particle
distribution with a `glass' configurations (White 1994). The time steps 
is 400 for SCDM evolving from $z=15$ to the present, and 600 for 
LCDM and OCDM from $z=20$.   

These models are degenerate at low redshift $z\sim$ 0, in the sense that
they produce almost the same abundance and power spectrum of clusters at
$z=0$. The degeneracies of SCDM - OCDM and SCDM - LCDM may be broken if the
abundance of clusters at high redshifts is considered. However, both LCDM
and OCDM predict almost the same redshift evolution of cluster abundance
until redshift $z\sim 0.5$.

\subsection{Scale-scale correlation of 3-D mass distribution}

To probe the scale-scale correlation, we placed randomly 1,000 cubic box of
$64^3$ $h^{-3}$ Mpc$^3$ in the $256^3h^{-3}$ Mpc$^{3}$ simulation space. 
These boxes form an ensemble of our statistics. Each $64^3$ box is 
divided into $128^3$ grids, and the $128^3$ 3-D matrix of density distribution 
is then obtained by the triangular 
shaped cloud (TSC) mass assignment scheme. We subject the 3-D matrix of density
field by a Daubechies 4 transform, and find the WFCs
$\tilde{\epsilon}_{\bf j,l}$, where $j=$1,2...6, corresponds to scales
$(64/2^j)$ $h^{-1}$ Mpc. 

The scale-scale correlation defined by eq.(6) is calculated for the 
ensemble of $64^3$ $h^{-3}$ Mpc$^3$ boxes in all the three cosmological
models. Figure 1 plots the histogram of the distribution of
$\log C_{j}^{2,2}$ for $j=2$ to 5 in the OCDM model. It shows that the
scale-scale correlation increases with the $j$, i.e. the scale-scale
correlation on smaller scale is greater than that on larger scales. The 
distribution
of $C_{j}^{2,2}$ at $j=2$, or $\sim \ 16$ $h^{-1}$ Mpc, appears to 
close a log-normal distribution. However, the mean value of
$C_{2}^{2,2}$ already departs from one. At $j=3$, i.e, at the spatial scale
$\sim \ 8$ $h^{-1}$ Mpc, the confidence level of $C_{3}^{2,2}$$>1$ is
$> 95 \%$. Down to the minimum sturcture resolved here, the distribution of 
$C_{5}^{2,2}$ is peaked as high as about
30, which indicates that the non-linear evolution of the density field
on small scales leads to a significant positive scale-scale correlation.

We now examine the model-dependence of the scale-scale correlation. 
Since the distributions of $\log C_{j}^{2,2}$ for $j \geq 3$ are highly 
non-Gaussian, the mean value and variance of $\log C_{j}^{2,2}$ is 
not enough to describe their statistical behavior. We introduce the cumulative 
distribution of $\log C_{j}^{2,2}$, defined by
\begin{equation}
F(<\log C_{j}^{2,2})=\frac{\int_0^{\log C_{j}^{2,2}} N(\log C_{j}^{2,2})d\log C_{j}^{2,2}}
{\int_0^{\infty} N(\log C_{j}^{2,2})d\log C_{j}^{2,2}}.
\end{equation} 
Figure 2 compares the cumulative distribution $F(<\log C_{5}^{2,2})$ of simulation samples in
the SCDM, OCDM and LCDM models at $z=0$. This figure shows clearly that the 
curve $F(<\log C_{5}^{2,2})$ of SCDM is significantly different from those of 
LCDM and OCDM. That is, low redshift ($z \sim 0$) data are already 
capable of breaking the degeneracy between SCDM and LCDM as well as OCDM if 
the scale-scale correlation of mass distribution could be detected. Moreover,
Figure 2 also shows that the curves $F(<\log C_{5}^{2,2})$ for LCDM and
OCDM are distinguishable from each other at low redshift, despite the 
difference between LCDM and OCDM is small relatively. This results, at least,
demonstrates the possibility of LCDM-OCDM degeneracy breaking by the DWT
scale-scale correlations.

The power of scale-scale correlation increases from SCDM to LCDM and
then OCDM. This is consistent with the sequence of non-linear evolution.
The SCDM clustering is less non-linear than LCDM and OCDM,  and the OCDM model
shows somewhat more compact structures than LCDM (Jing et al. 1995).

\subsection{Scale-scale correlation of 2-D mass distributions}

To measure the scale-scale correlation in 2-D distribution, we make an 
ensemble of samples by randomly placing 1,000 $128\times128\times16$ 
$h^{-3}$ Mpc$^3$ slabs in the simulation space. A mass distribution
in 2-D space of $128\times128$ ($h^{-1}$ Mpc)$^2$ can be obtained by 
summing up all masses along the dimension of 16 $h^{-1}$ Mpc. The density 
distribution is tabulated by TSC mass assignment scheme on $1024^2$ grids 
in the 2-D. Accordingly, the $j$ runs 1,2,...,9, which corresponds to 
scales $(128/2^j)$ $h^{-1}$ Mpc.  

Figure 3 presents the histogram of the $\log C_{j}^{2,2}$ distribution for the
2-D samples in the OCDM model. Obviously, the figure shows the similar
behavior as 3-D scale-scale correlation. The power of the 2-D
scale-scale correlations $C_{j}^{2,2}$ is almost the same as that for the
3-D samples but with the slightly smaller values than in the 3-D case.
Generally, the ordinary two-point correlation strength in 2-D is 
significantly smaller than in 3-D due to the projection effect. However, the 
$C_{j}^{2,2}$ measures an ratio between localized correlations, the effect 
of projection is weaker. Therefore, detecting scale-scale correlations in
2-D samples would be as effective as in 3-D cases.

The model-dependence of $C_{j}^{2,2}$ in the 2-D samples are shown in 
Figure 4, which compares the cumulative distribution $F(<\log C_{7}^{2,2})$ between
the SCDM, OCDM and LCDM models at $z=0$. Similar to the 3-D case, the curve
of $F(<\log C_{7}^{2,2})$ for the SCDM is substantially different from that of
the LCDM and OCDM models. The curves for the LCDM and the OCDM are also 
distinguishable. This result does not very sensitively depend upon the
thickness of the slab. Using the slab of
$128\times128\times32$ $h^{-3}$ Mpc$^3$, we found very similar results.
Thus, to break the degeneracy between the SCDM, LCDM and OCDM models, 
the 2-D data at low redshift ($z \sim 0$) are as powerful as 3-D samples.

\subsection{Scale-scale correlation of biased galaxy distributions}

Now we turn to the scale-scale correlation of simulation samples of galaxies.
In order to produce galaxy samples comparable with the mock samples of
SDSS and 2dF galaxy redshift surveys in Cole et al. (1998), we employ the 
so-called two-parameter Lagrangian bias model to identify the galaxies from 
N-body simulation mass distributions. Namely, the density field is first 
smoothed by a Gaussian window function of width 3$h^{-1}$ Mpc. The biased 
galaxies distribution is then drawn from the smoothed density field 
$\rho({\bf r})$ by a selection probability defined as
\begin{equation}
P(\nu) \propto \left \{ \begin{array}{ll}
      \exp(\alpha\nu+\beta\nu^{3/2}) & \mbox{ if $\nu\ge 0$}, \\ \nonumber
      \exp(\alpha\nu) &   \mbox{ if $\nu\le 0$}
      \end{array}
\right. 
\end{equation}
where the dimensionless variable $\nu$ is defined by
$\nu({\bf r})=\delta({\bf r})/\sigma$; 
$\delta({\bf r})=(\rho({\bf r})-\bar{\rho})/\bar{\rho}$ the density contrast
of the density field $\rho({\bf r})$, and
$\sigma^2=\langle|\delta_S^2|\rangle$ its variance.
The two parameters $\alpha$ and $\beta$ are determined by the least
square-fitting of the observed variance of galaxies count in cubic cell of
$5$ $h^{-1}$ Mpc and $20$ $h^{-1}$ Mpc. The adopted values $\sigma_5=2.0$ and
$\sigma_{20}=0.67$ were obtained for APM galaxies samples by Baugh and
Efstathiou (1994). This two parameter bias models are more flexible in
matching both of the amplitude and the slope of the galaxy two-point
correlation function on scales $1-10$ $h^{-1}$ Mpc, though their underlying
matter distribution may be quite different. That is, all models used to
producing the biased galaxy samples are degenerate in terms of the low 
order statistics such as two-point correlation or power spectrum of galaxies.

Proceeding in same way as used in previous subsection, we placed randomly 1,000
cubic boxes of $64^3$ $h^{-3}$ Mpc$^3$ in the simulation space. The biased
galaxy samples from these $64^3$ boxes form an ensemble for our statistics
of the scale-scale correlations. The distributions of $C_{j}^{2,2}$ for
the OCDM model are presented in Figure 5.  The $C_j^{2,2}$ shows the similar
behavior of scale-dependence as the mass distribution, i.e. the scale-scale
correlation increases with the $j$. However, the values of
$C_{j}^{2,2}$ are less than their parent mass distributions.

We also produced biased galaxy samples in redshift space. For simplicity,
the redshift distortion is applied along one axis of 3D Cartesian
coordinates, which is in fact equivalent to the assumption of an infinitely
distant observer. The 1000 randomly placed $64^3$ $h^{-3}$ Mpc$^3$ boxes 
form an ensemble for calculating scale-scale correlations. The results are 
also shown in Figure  5. The scale-scale correlations in redshift space are 
generally less than those in real space, i.e, redshift distortion suppresses 
the scale-scale correlation.  The difference of the scale-scale correlation 
between the redshift and real spaces disappears gradually with increasing
scales.

This result is just what we expect. The local scale-scale correlations
are sensitive to the accuracy of distance or position measurement. A
radial distance uncertainty in redshift space is caused by the peculiar
velocity. Therefore, the scale-scale correlation will be significantly
contaminated on scales comparable with the uncertainty of radial distance,
while the correlation on large scale will not be affected by the redshift
distortion.

To reduce the effect of redshift distortion, one can measure the scale-scale
correlation in 2-D samples, as 2-D sample is not affected by redshift
distortion. As shown in \S 3.3, the power of the scale-scale correlation of
2-D mass distribution is about the same as 3-D sample in real space.
Therefore, 2-D sample probably is more suitable than 3-D samples to detect 
scale-scale correlations on scales where the redshift distortion is 
significant.

\subsection{Non-linearity of bias}

An advantage of the statistics $C_j^{p,p}$ is free of linear-bias
parameters. Linear bias parameter $b$ is defined by
\begin{equation}
\delta({\bf x})_{g} = b \delta({\bf x})_{m},
\end{equation}
where $\delta({\bf x})_{g}$ and $\delta({\bf x})_{m}$ are 
the density contrasts of galaxies and density dark matter, respectively.
Since the DWT is linear, one has 
\begin{equation}
\tilde{\epsilon}_{\bf j,l}\ _{g} = b
\tilde{\epsilon}_{\bf j,l}\ _{m}.
\end{equation}
Therefore, we have 
\begin{equation}
C_{\bf j}^{p,p}\ _{g}=C_{\bf j}^{p,p}\ _{m}
\end{equation}
If the linear bias is introduced by the variance of perturbations, i.e.
\begin{equation}
\tilde{\epsilon}^2_{\bf j,l}\ _{g} = b
\tilde{\epsilon}^2_{\bf j,l}\ _{m},
\end{equation}
we still have eq.(12).

Accordingly, one can detect the non-linearity of a bias model
by comparing the $C_{\bf j}^{2,2}$ of the biased galaxy samples
with the underlying unbiased matter distribution. As have been seen
from Figures 1 and 5, the values of $C_{j}^{2,2}$ of the biased galaxies
are significantly less than their parent mass distributions.
Therefore, the bias model eq.(9) is highly non-linear.

Figure 6 gives the spectra of $C_j^{2,2}$ for both biased and
unbiased galaxy samples in the OCDM model. Obviously, the unbiased sample
has much stronger scale-scale correlation than biased galaxies created
by eq.(9).

\section{Scale-scale correlation of APM-BGC galaxies}

\subsection{Samples of APM-BGC galaxies}

Now, we attempt to measure the scale-scale correlation in the observational 
galaxy samples, and then make comparison with the model predictions. We 
analyzed the 2-D samples of galaxies listed in the APM bright galaxies 
catalog (Loveday et al. 1996), which gives positions, magnitudes
and morphological types of 14,681 galaxies brighter than 16$^m$.44 over a
4,180 deg$^{\circ}$ area in 180 Schmidt survey fields of south sky. The
completeness is about 96.3\% with a standard deviation 1.9\% inferred
from carefully checked 12 fields. Therefore, it is large and uniform enough
for a 2-D DWT analysis.

In order to carry out a 2-D DWT analysis, we chose two fields S1 and S2 from 
the entire survey area as shown in Figure 1 of Fang, Deng \& Xia (1998). S1 
and S2 do not overlap from each other. These two fields are selected to be 
squares on an equal area projection of the sky. The equal area projection 
keeps the surface number density of galaxies on the plane being the same as 
that on the sky. The sizes of S1 and S2 have been taken to be as large as 
possible to cover the whole area of the survey. Each field has angular size 
of about 37$^{\circ}\times37^{\circ}$. S1 and S2 contain 1,095
and 1,055 elliptical and lenticular (EL) galaxies, and 2,186 and 2,092
spiral (SP) galaxies, respectively. The galaxies in the regions S1 and S2
are completely independent.

We divide each square region into 2$^{10}$ $\times$ 2$^{10}$ (1024$^2$)
cells, and the angular scale labeled by $j$ is $2^{10-j}$ in unit of the size
of a cell, $j$ runs from 1...10. Using the estimation of the mean depth of
the sample given by the luminosity function from the Stromlo-APM redshift
survey (Loveday et al. 1992), the cell size is found to be about
65 h$^{-1}$Kpc, which is fine enough to detect statistical features on
scales larger than 1 h$^{-1}$Mpc.

In doing statistical analysis of the APM-BGC, the effect of the 1,456 holes 
drilled around big bright objects should be taken into account. For instance, 
random samples are generated in the area with the same drilled holes 
as that in the original survey. This may lead to a underestimation of the
scale-scale correlations. We also removed a few galaxies placed in the 
holes drilled.

\subsection{Sales-scale correlations of EL and SP galaxies}

Figure 7 plots the $C_{j}^{2,2}$ of the EL and SP galaxies in the fields I
and II. The confidence level is estimated from 1,000 randomized samples
with the same number of galaxies as real data. Both randomized and bootstrap
resampling give about the same confidence level (Fang, Deng \& Xia 1998). 
Figure 6 shows that the randomized distributions are also substantially
scale-scale correlated for $j =$ 5, 6, i.e. $C_{5}^{2,2}, \ C_{6}^{2,2} > 1$.
This is because the random sample on small scales ($j =$ 5, 6) essentially is
a Poisson process which is non-Gaussian and thus leads to scale-scale
correlation. On larger scales ($j \leq$ 4), the mean number of galaxies in
a $j$ cell is high enough, and their one-point distribution of random
samples is Gaussian.

Despite the shot noise, Figure 7 shows that the distributions of the EL and 
SP galaxies of the APM bright galaxies catalog are highly scale-scale 
correlated at $j=5, \ 6$. The scale-scale correlations are also significant 
($>95\%$ confidence level) on scales $j=3$ and 4, which corresponds to the 
scale $\sim 8$ h$^{-1}$ Mpc. The EL and SP galaxies basically have about the 
same behavior of the $j$-dependence of $C_{j}^{2,2}$. It increases with the 
increase of $j$. This is qualitatively consistent with our simulation results 
(\S 3). In addition, ELs generally have higher power of $C_{j}^{2,2}$ than 
SPs. This is consistent with the segregation bewteen ELs and SPs. 
ELs concentrate in clusters, having higher clustering, while the field
galaxy SPs are lower.

\subsection{Mock samples of galaxies }

For a quantitative confrontation between observed and model-predicted 
scale-scale correlations, we use the mock samples for the SDSS redshift
survey (Cole et al. 1998). The radial selection effect is prescribed by the 
Schechter form of the luminosity function with the parameters drawn from the 
the Stromlo-APM bright galaxy survey (Loveday, et.al, 1992). Therefore, it is 
appropriate to use these mock samples for demonstrating the
model-degeneracy breaking by the APM-BGC galaxy scale-scale correlations.

We consider still the three popular cosmological models of SCDM, LCDM and 
OCDM here (referred as E4, L3S and O3S respectively in Cole et al, 1998). 
To mimic APM-BGC sample, we extract 10 subsamples out of the SDSS 
mock catalogue by imposing magnitude limit of $B_j\le 16.44$ for each model of 
the SCDM, LCDM and OCDM. The size of the 10 fields is the same as the fields
I and II. The 10 fields are chose to cover the projected survey area as large 
as possible but have less overlap from each others. This selection ensures 
the statistcal independence to a maximum extent.   

The surface number density of galaxies for all the mock samples is found to
be approximately the same as that obtained for APM-BGC data, the difference 
between the number densities of the real data and mock samples is less than 
3\%. The two-point angular correlation functions of the mock samples in the 
three cosmological models are also found to be in good agreement with that of 
APM-BGC. That is, in terms of the first (number density) and the second order
(two-point correlation function) statistics, the mock samples of the
SCDM, LCDM and OCDM are degenerate. 

\subsection{The APM-BGC scale-scale correlation and cosmological models}

Since the mock samples do not contain the information of morphology, we
calculated the scale-scale correlations of APM-BGC galaxies consisting of
both ELs and SPs in the fields I and II. The results are shown in Figure 8.
The 95\% confidence level is estimated from 1,000 Poisson-distributed samples
with the same number of galaxies as the real data. The scale-scale correlations
of the 10 mock samples in the SCDM, LCDM and OCDM models are displayed by the 
scatter symbols in Figure 8 repectively.

Figure 8 shows first that the fields I and II give almost the same spectrum
of $C_{j}^{2,2}$ vs. $j$. The data are qualified for model discrimination.
At the first glance, all mock samples of the three CDM cosmogony models
show the same shape of the $C_{j}^{2,2}$ spectrum, and are generally
consistent with the observed $C_{j}^{2,2}$. However, SCDM, LCDM and OCDM
predicted different $C_{j}^{2,2}$ at $j=5$. The OCDM is in good agreement
with the entire spectrum of APM-BGC's $C_{j}^{2,2}$ from $j=2$ to 5, i.e.
scales $\sim 2$ to 16 h$^{-1}$ Mpc. However, the values of $C_{5}^{2,2}$ of
SCDM and LCDM are somewhat lower than the observation at $90\%$ confidence
level. Considering the possible underestimate of the APM-BGC's scale scale
correlation due to the holes drilled around big bright objects, the difference
between observed results and the SCDM and LCDM prediction might be more 
significant than $90\%$. Thus, the test of scale-scale correlation seems to 
be in favor of the OCDM model.

It might be too early to say this conclusion is concrete, as the data is
still not perfect enough. However, this result showed that the scale-scale
correlation is effective and feasible as a tool of degeneracy breaking. The
correlations are able to reveal not only the difference between the SCDM
and OCDM, but also the LCDM and OCDM.

All $C_{j}^{2,2}$ shown in Figs. 1 - 8 are calculated by taking
$\Delta {\bf l} =0$ [see eq.(7)]. It describes the correlations between
$j$ and $j+1$ perturbations at the same physical coordinates, i.e. it is
local scale-scale correlation. We also calculated non-local scale-scale
correlations by taking $\Delta l = 1, \ 2,$ and 3. It describes the
correlation between perturbations on scale $j$ at position $l$ and on
scale $j+1$ at $l\pm \Delta l$. Figure 9 compares the non-local correlations
$C_{j, \Delta l=1, 2, 3}^{2,2}$ of APM-BGC data and the OCDM mock sample.
It shows that the OCDM result is in excellent agreement with observed
data.

\section{Discussions and conclusions}

The ordinary two-point correlation function is two-point statistics in
physical space. The scale-scale correlations essentially are two-point
statistics both in physical space and wavenumber space, i.e., in 
a position-scale phase space. Moreover, the scale-scale correlation describes
the feature of the non-linear evolutions of the hierarchical clustering.
Therefore, the scale-scale correlation is probably the most basic measure
of the cosmic clustering in the phase space.

The abundance and two-point correlation functions of galaxies and clusters
are unlikely to discriminate among the CDM family of models, which are
degenerate in cosmological parameter space. In this paper, using the mass
distribution generated by N-body simulation, we showed that the scale-scale
correlation of mass distributions is sensitive to the cosmological
parameters. The scale-scale correlation of biased galaxy distribution is
sensitive to cosmological and biasing parameters. Therefore, scale-scale
correlation of galaxy distributions provides us a tool of discriminating
among various cosmological models, which are degenerate in the parameter
space, including both cosmological and bias parameters.

Using APM-BGC data, we found that the $j$ spectrum of local and non-local
scale-scale correlation can generally be matched by the mock galaxy samples
of the CDM cosmogonies. Amongst the models, the best fitting is given by
the OCDM model. Both of SCDM model and LCDM model show somewhat weaker
scale-scale correlations than OCDM on small scales. We can conclude
that the OCDM plus the biasing algorithms (9) is favored by the current
detection of APM-BGC's scale-scale correlations. This showed that the
scale-scale correlation is a capable tool of degeneracy breaking.

We found that the scale-scale correlation in 2-D samples is comparable with 
that in 3-D. Because the redshift distortion contaminates the scale-scale 
correlation in 3-D, and meanwhile the 2-D samples is free of redshift 
distortion, the scale-scale correlation measured in 3-D redshift galaxy 
survey will be lower than their 2-D scale-scale correlation. Therefore, the 
difference between 2-D and 3-D may provide us a new way to study
the large scale velocity field.

We thank Dr. Wolung Lee for useful discussion. LLF acknowledge support from the 
National Science Foundation of China(NSFC) and World Laboratory scholarship. 
ZGD is supported by NSFC and exchange program between NSFC and DFG. This project was 
done during LLF's visiting at Department of Physics, University of Arizona.

\clearpage

\figcaption{The histogram of the scale-scale correlations, $C_{j}^{2,2}$,
of mass distribution for OCDM model. $C_{j}^{2,2}$ is computed for 1000 
randomly throwing box of 64 $h^{-1}$ Mpc on a side.}
\label{fig1} 
 
\figcaption{The cumulative distribution of $\log C_{5}^{2,2}$ of mass 
distributions for models of SCDM, OCDM and LCDM.}
\label{Fig2}

\figcaption{The histogram of the scale-scale correlations, $C_{j}^{2,2}$,
of 2-D mass distribution for OCDM model. $C_{j}^{2,2}$ is computed for 
1000 randomly throwing box of 64 $h^{-1}$ Mpc on a side.}
\label{fig3} 
 
\figcaption{The cumulative distribution of $\log C_{7}^{2,2}$ of 2-D mass
distributions for models of SCDM, OCDM and LCDM. For the same type of line, 
the left and right corresponds to the slabs of thinkness 32 $h^{-1}$ Mpc 
and 16 $h^{-1}$ Mpc, respectively.}
\label{Fig4}

\figcaption{The histogram distribution of the scale-scale correlations, 
$C_{j}^{2,2}$, of OCDM biased galaxy distributions in both real space
and redshift space (filled histogram).}
\label{Fig5}

\figcaption{The spectrum of 3-D scale-scale correlation in the range with 90\% confidence
level both for the underlying matter distribution (dark gray band) and biased
galaxies (light gray band) in OCDM model. The solid lines are given by
ensemble average of $\log C_j^{2,2}$. In addition, the scale-scale correlation
for Poisson-distributed galaxy samples(black band) is also plotted.
}
\label{Fig6}

\figcaption{The 2-D local ($\Delta l =0$) scale-scale correlation of EL
and SP samples in areas I and II of APM-BGC survey. Gray band shows the
range with 95\% confidence level for 1,000 Poisson-distributed galaxy samples.
}
\label{Fig7}

\figcaption{The 2-D local ($\Delta l =0$) scale-scale correlation of ELs
plus SPs in areas I and II of APM-BGC galaxy sample (solid lines).
The square represents the value obtained from one of the 10 mock samples 
of the indicated model (SCDM, OCDM and LCDM) and are displayed in the range with 90\% 
confidence level. The gray bands show the range with 
95\% confidence level for Poisson-distributed galaxy samples.
}
\label{Fig8}

\figcaption{The 2-D non-local scale-scale correlation with
$\Delta l =1, 2,$ and 3 in OCDM. The solid line, squares and gray band 
are the same meaning as those in Fig. 8.
}
\label{Fig9}

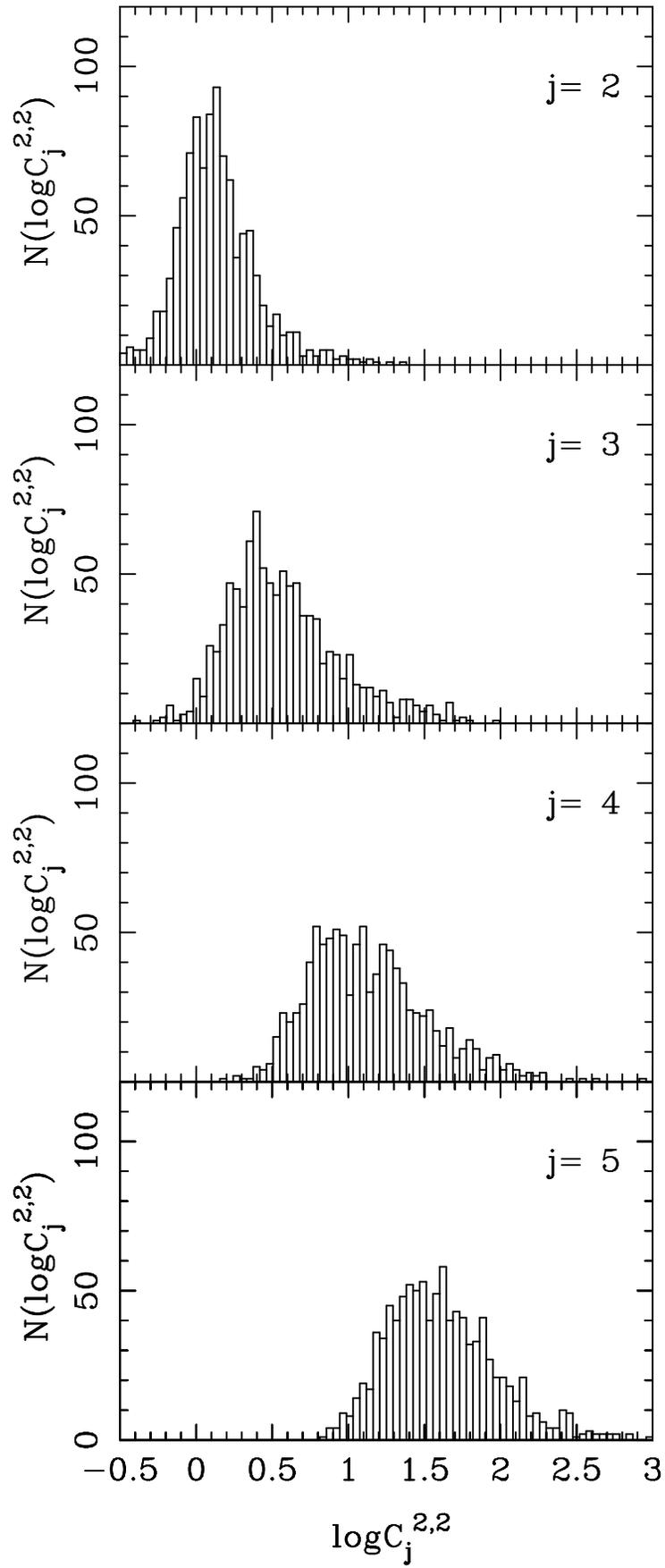
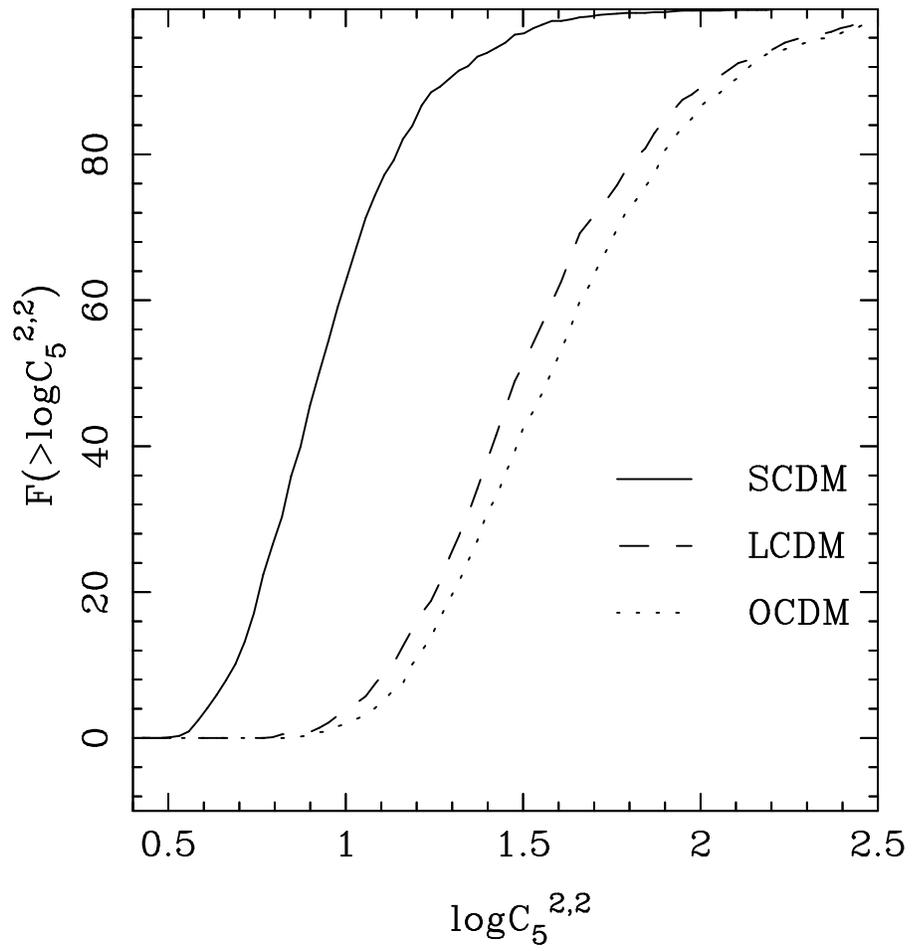
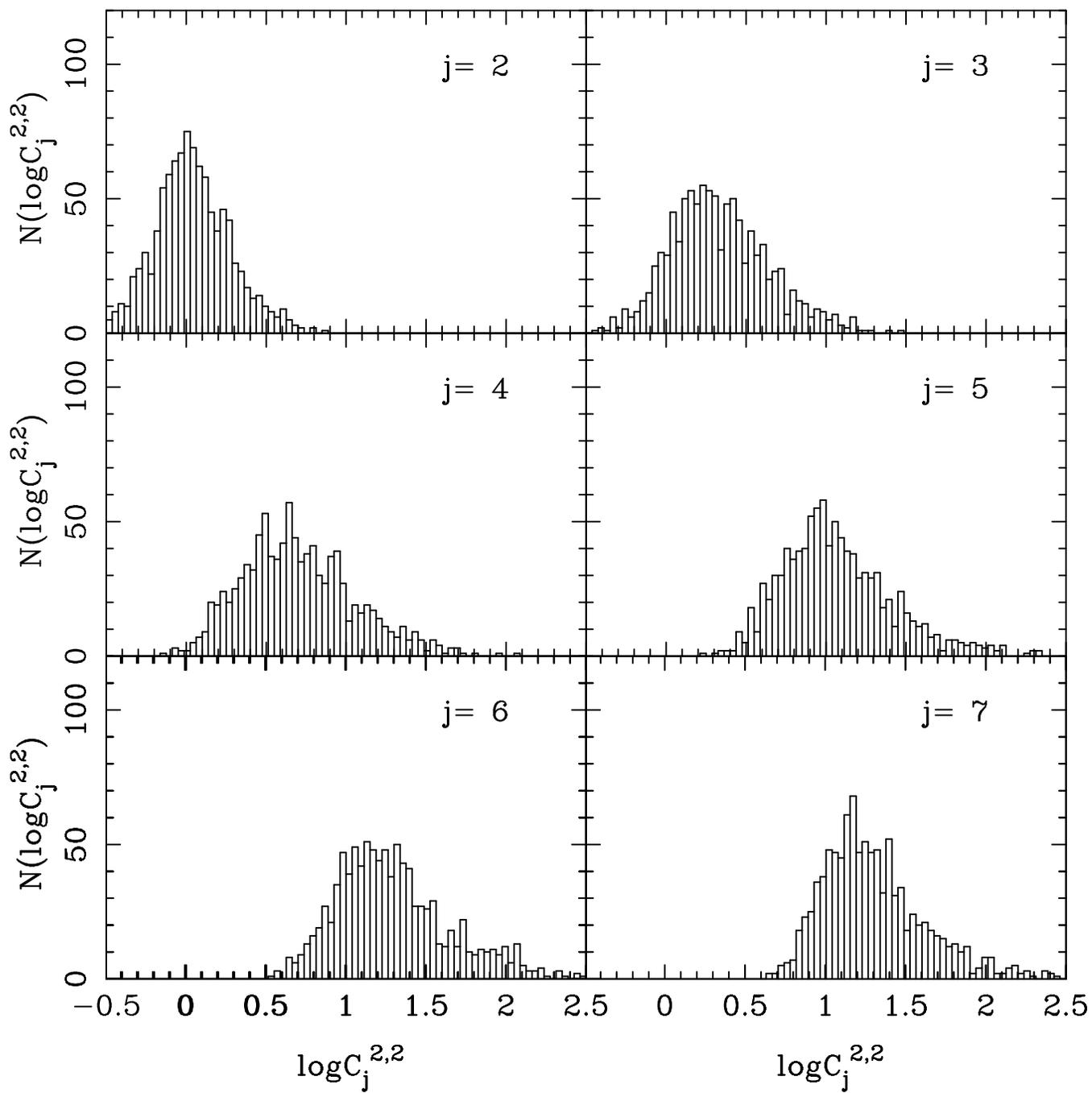
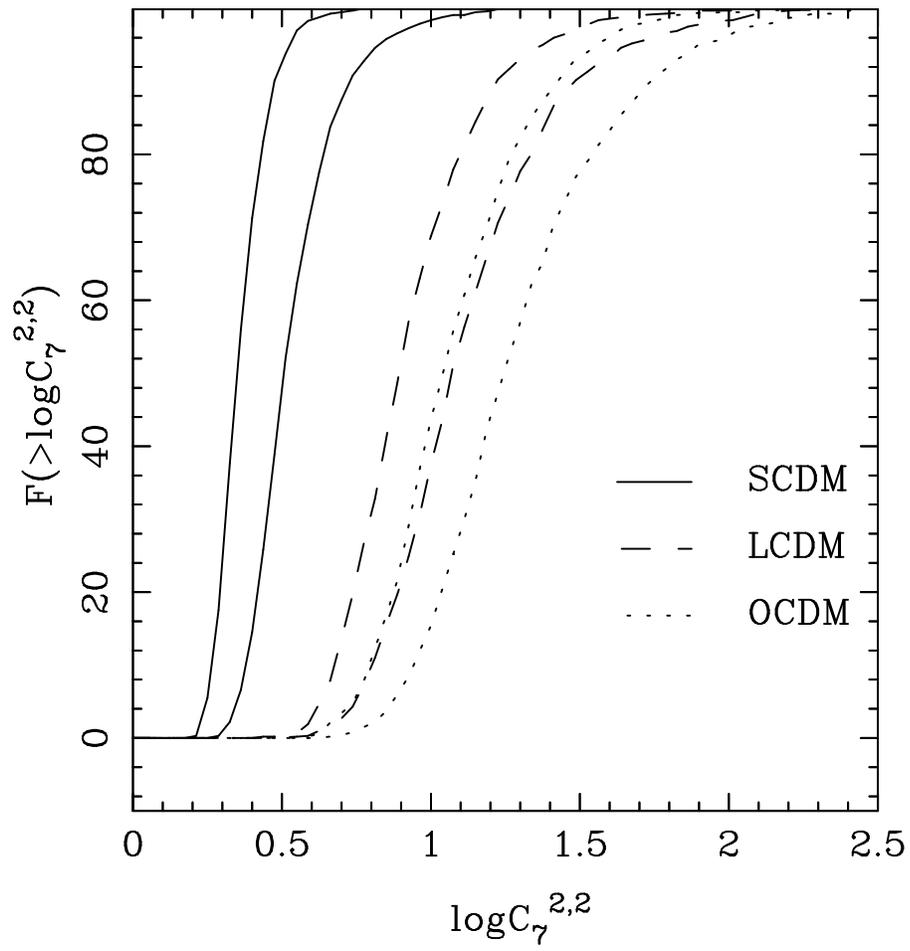
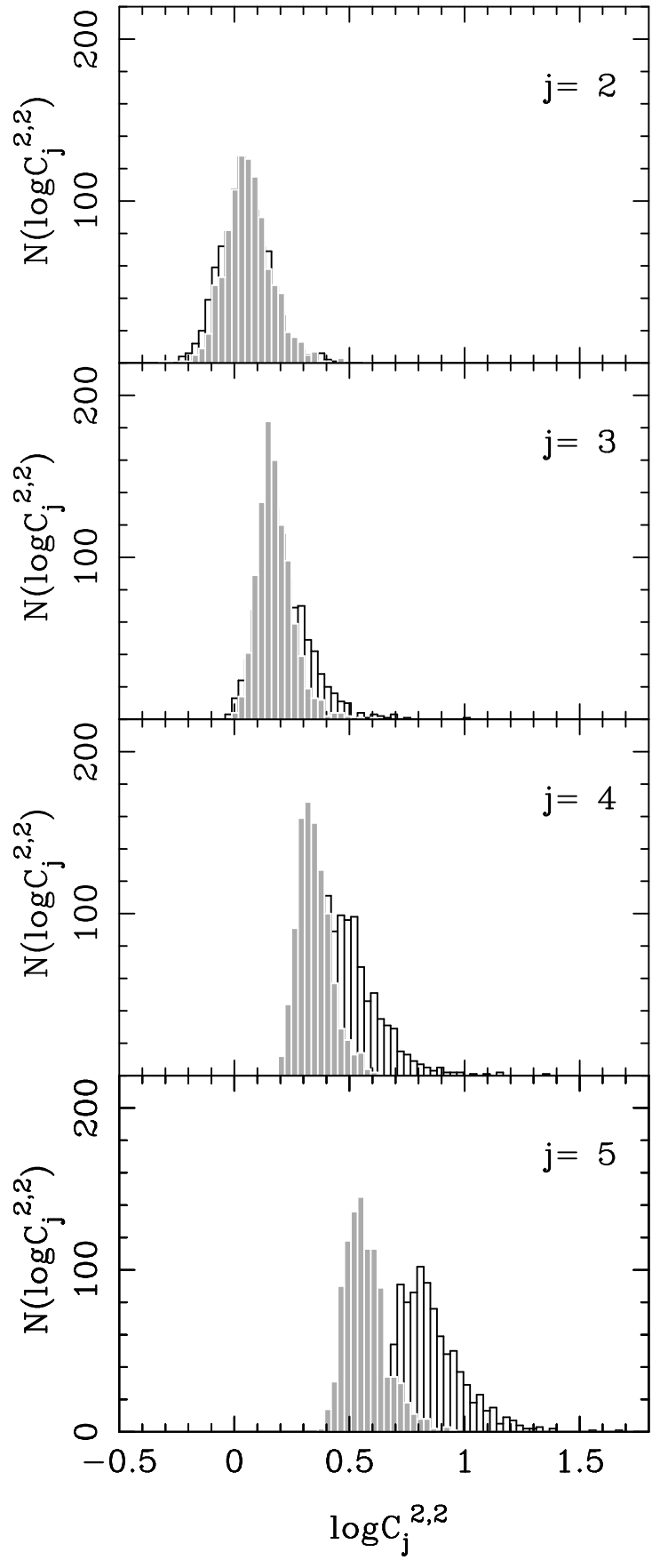
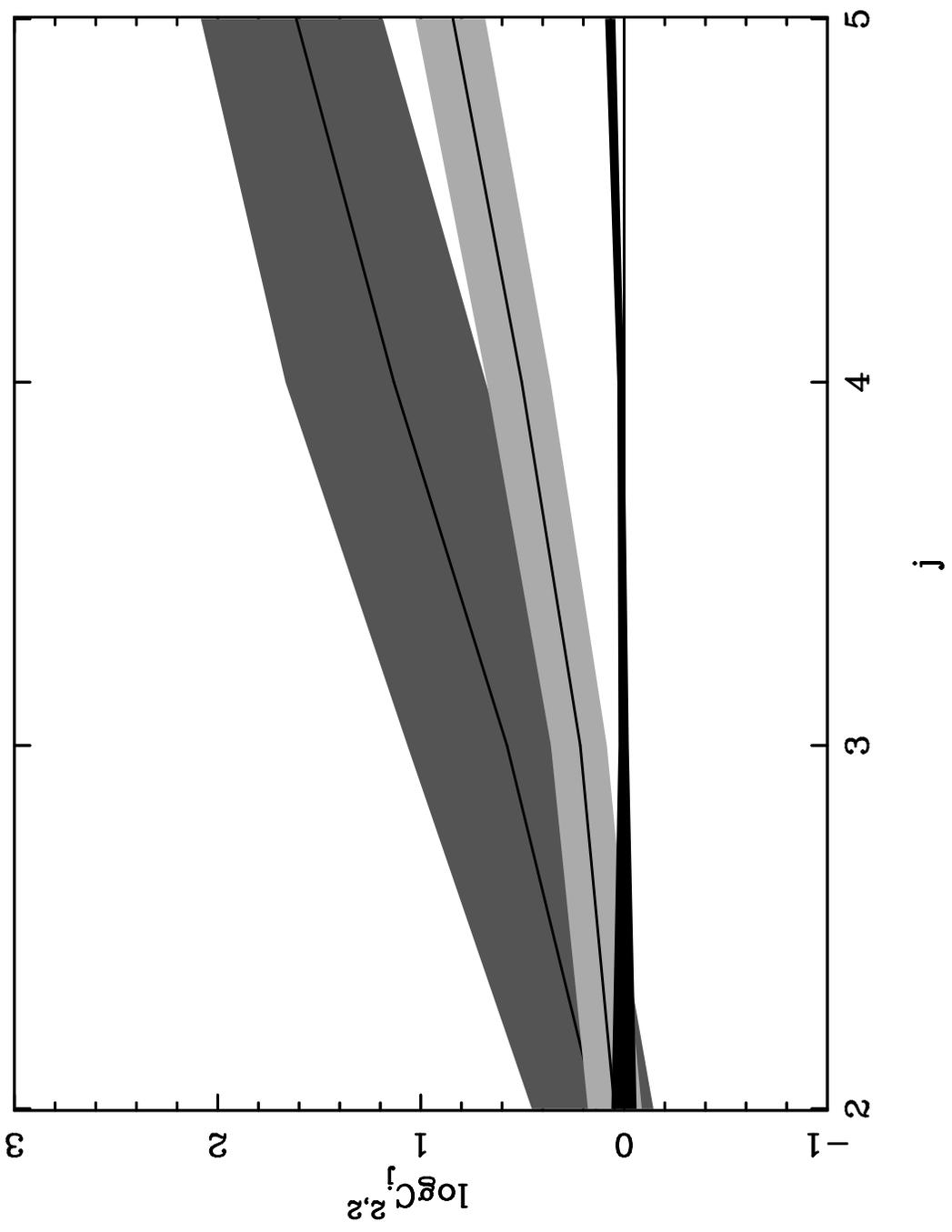
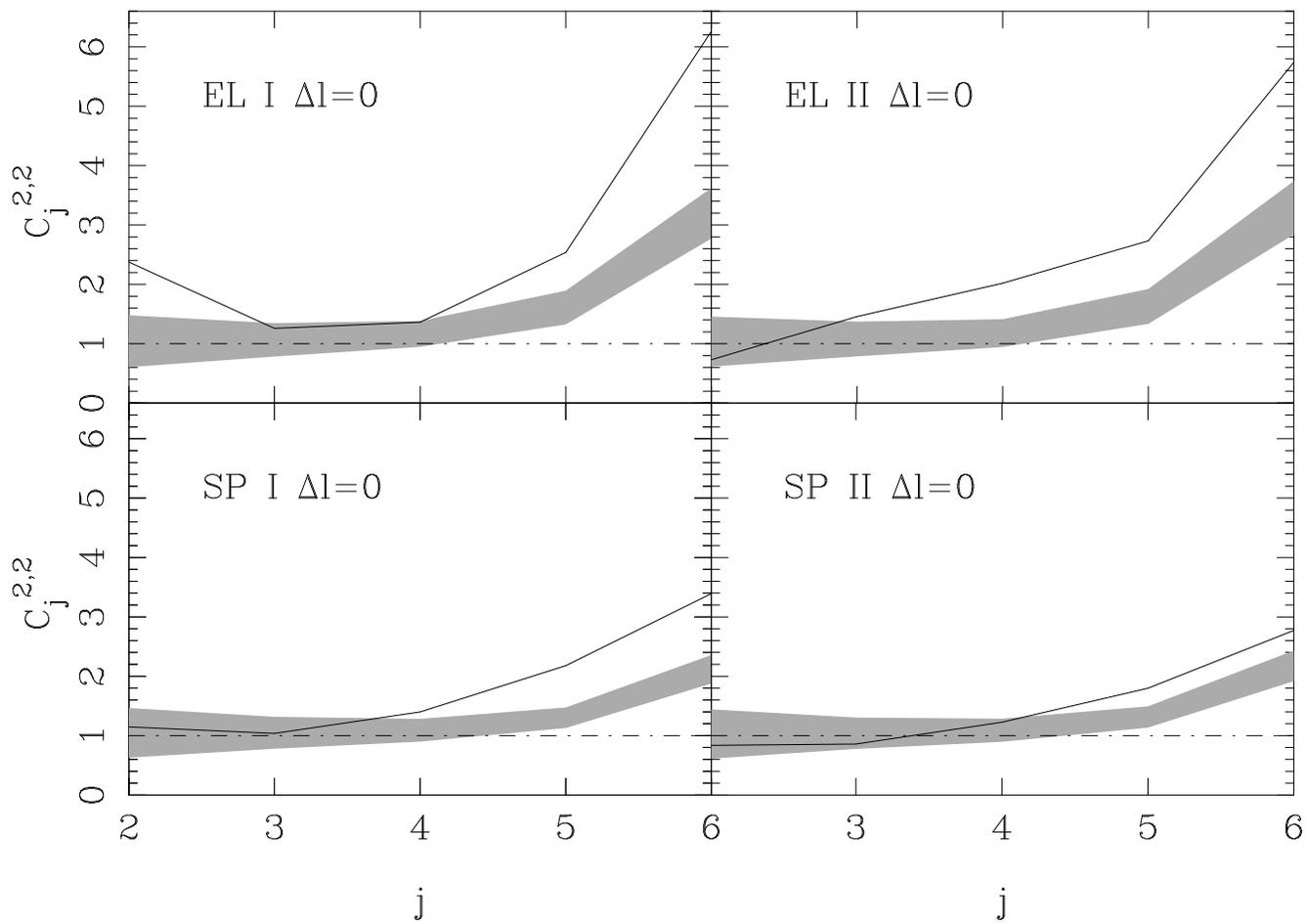
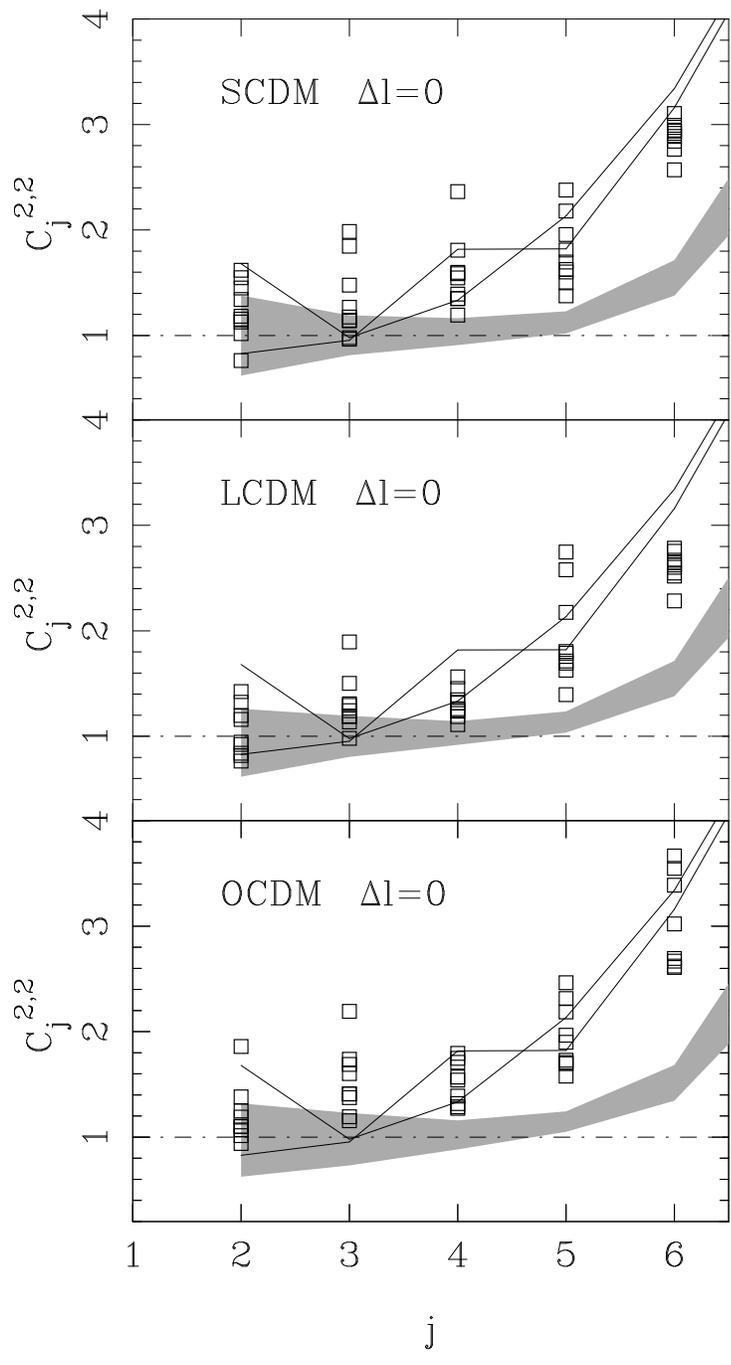
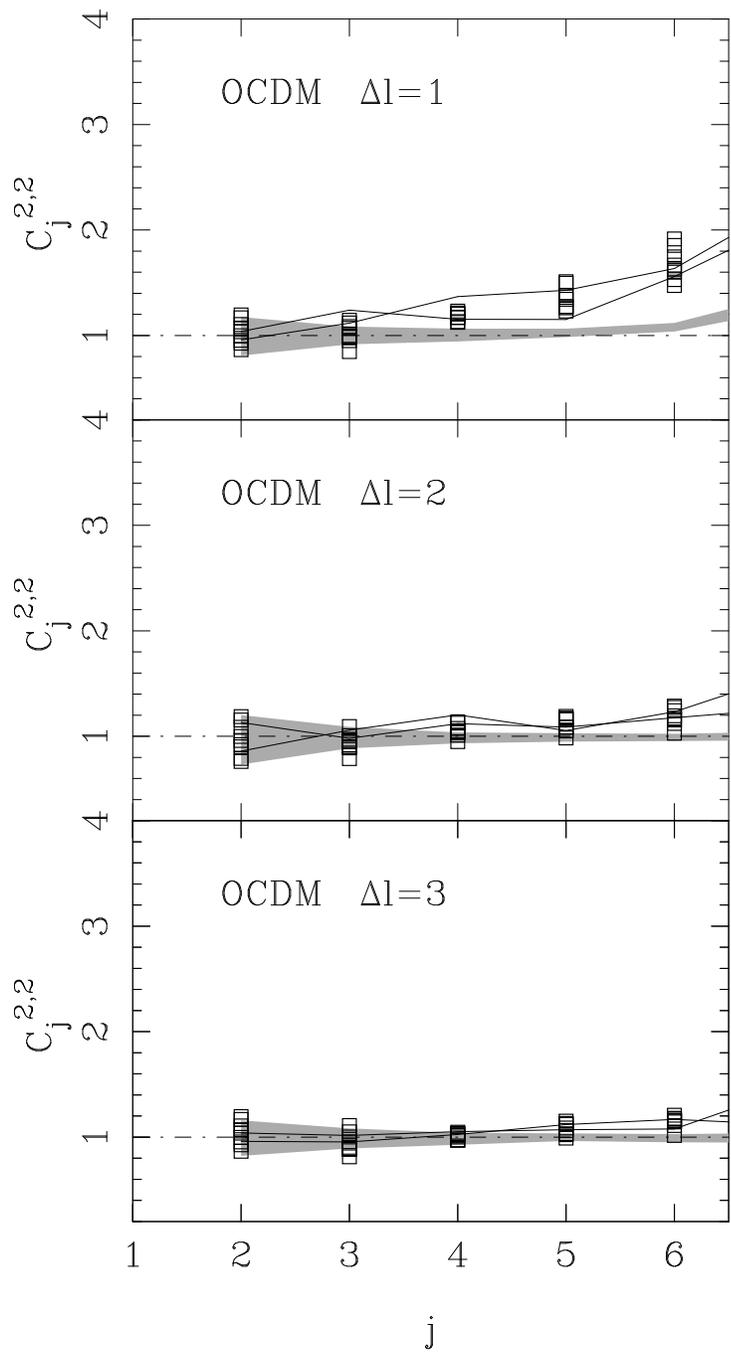
\end{document}